\documentclass[reqno]{article}
\usepackage{alltt}
\usepackage{graphicx}
\title{Quantum geometry of the Cartan control problem}
\author{Peter Leifer}
\date{Cathedra of Informatics, Crimea State Engineering and
Pedagogical University, \\
21 Sevastopolskaya st., 95015 Simferopol, Crimea, Ukraine; \\Hermon Laboratories, Ltd. \\
Binyamina, 30500 Israel \\
leifer@bezeqint.net, peter@hermonlabs.com }
\begin{document}
\maketitle
\begin{abstract}
The Cartan control problem of the quantum circuits discussed from the differential geometry point of view. Abstract unitary transformations of $SU(2^n)$ are realized physically in the projective Hilbert state space $CP(2^n-1)$ of the n-qubit system. Therefore the Cartan decomposition of the algebra $AlgSU(2^n-1)$  into orthogonal subspaces $h$ and $b$ such that $[h,h] \subseteq h, [b,b] \subseteq h, [b,h] \subseteq b$ is state-dependent and thus requires the representation in the local coordinates.
\end{abstract}


\section{Introduction}
The optimal quantum state evolution is one of the most important problem in quantum
computations \cite{GDN}. Physically it comprises the minimization of the time to synthesize desirable Hamiltonian capable unitary connect initial and target quantum states.
The important achievement on this way is the understanding the invariant (geometry) properties of the unitary group \cite{KN}. I mean the factorization of unitary transformation according to the Cartan's decomposition of the algebra $AlgSU(N)$.

If we take traceless hermitian matrices like Pauli, Gell-Mann one sees that they may be divided in two subsets $h$ and $b$ such that $[h,h] \subseteq h, [b,b] \subseteq h, [b,h] \subseteq b$.
The $h$ direction in algebra correspond to isotropy subgroup of some state vector, i.e.
unitary transformations that leave this state vector intact. The $b$ direction correspond coset transformations which deform the chosen state vector. For each given $N=2^n$ one could build $N^2-1=2^{2n}-1$ traceless ``basis" matrices that may be divided in two sub-sets
$h$ and $b$. One should remember, however, that the Cartan decomposition of interaction Hamiltonians treated as ``a physical resource" \cite{GDN} has the physical sense only in respect with initially chosen state vector. Therefore the parametrization of these decomposition is state-dependent $[h_{\Psi},h_{\Psi}] \subseteq h_{\Psi}, [b_{\Psi},b_{\Psi}] \subseteq h_{\Psi}, [b_{\Psi},h_{\Psi}] \subseteq b_{\Psi}$ \cite{Le1,Le2,Le3}. It means that physically it is interesting not abstract unitary group relations but realization of the unitary group transformations resulting in motion of the pure quantum states represented by rays in projective Hilbert space.
\section{The Hamiltonian control problem for initial and target states}
The Hamiltonian control problem was initially formulated by M.A. Nielsen \cite{MAN}.
Recently it it was reformulated to the ``quantum control problem" \cite{GDN} as synthesis
of given n-qubit unitary $U \in SU(2^n)$ by the application of some time dependent Hamiltonian $H(t) \in AlgSU(2^n)$ during time $T$ so that the
total cost $D(U,H)=\int_0^T C(U,H) dt$ is minimized with so-called cost function $C(U,H)$.
This cost function was chosen with help ``Cartan control problem". I however propose to simplify this problem even more so that in this framework  ``the physical interpretation of n-qubit Cartan control problem for $n > 2$ is not as transparent" \cite{GDN}.
Hereafter I will use for simplicity the symbol $N$ assuming $N=2^n$ where it is necessary.

Taking into account that only quantum states have physical sense I propose to estimate
``the cost" as a geodesic distance between initial and target quantum states in the projective Hilbert space $CP(N-1)$ endowed by Fubini-Study metric \cite{KN}.
In fact it is possible and reasonable because the cost of isotropy group $H=U(1) \times U(N-1)$ action is not simply negligible in comparison with the coset $CP(N-1)=S[U(N)/U(1) \times U(N-1)]$ action but identically equal to zero. Isotropy group of the initially chosen vector acts only as gauge transformations whereas the coset transformations represent ``quantum force" deforming quantum state \cite{Le1}.

\textbf{Theorem 1.} For arbitrary $N$ the ``squeezing" unitary transformation $U \in SU(N)$  may be represented by $N-1$ quantum gates acting step-by-step on $N-1$ qubits.

\textbf{Proof.} Let me assume that we have two state vectors $|A>, |B> \in C^N$.
Applying step-by-step the ``squeezing ansatz" \cite{Le2}, say to initial vector
\begin{eqnarray}\label{1}
|A>= \left( \matrix {a^1 \cr a^2 \cr a^3 \cr . \cr . \cr . \cr a^N }
\right).
\end{eqnarray} one may reduce it to the form

\begin{eqnarray}\label{2}
z|1>= e^{i \phi} ||A|| \left( \matrix {1 \cr 0 \cr 0 \cr . \cr . \cr . \cr 0 }
\right).
\end{eqnarray}

I will apply now the ``squeezing ansatz" to vector $|A>$.
The first ``squeezing'' unitary matrix is
\begin{equation}\label{3}
\hat{U}_1= \left( \matrix{1&0&0&.&.&.&0 \cr 0&1&0&.&.&.&0 \cr
.&.&.&.&.&.&. \cr .&.&.&.&.&.&. \cr 0&.&.&.&1&0&0 \cr .&.&.&.&0&\cos
\phi_1&e^{i\psi_1} \sin \phi_1 \cr 0&0&.&.&0&-e^{-i\psi_1} \sin
\phi_1&\cos \phi_1 } \right ).
\end{equation}
This transformation being applied to our $|A>$
with the following result
\begin{eqnarray}\label{4}
|A>= \left( \matrix {a^1 \cr a^2 \cr a^3 \cr . \cr . \cr a^{N-1} \cos \phi_1 +a^N e^{i\psi_1} \sin \phi_1  \cr -a^{N-1}e^{-i\psi_1} \sin
\phi_1 + a^N \cos \phi_1  }
\right).
\end{eqnarray}
Now one should solve two ``equations of annihilation'' of $\Re (-a^{N-1}e^{-i\psi_1} \sin
\phi_1 + a^N \cos \phi_1 )=0$ and $\Im (-a^{N-1}e^{-i\psi_1} \sin
\phi_1 + a^N \cos \phi_1 )=0$ in order to eliminate the
the last row. This gives us $\phi'_1$ and $\psi'_1$.
The next step is the similar transformation given by the unitary matrix with the
diagonally shifted up the transformation block
\begin{equation}\label{5}
\hat{U}_2= \left( \matrix{ 1&0&0&.&.&.&0 \cr 0&1&0&.&.&.&0 \cr
.&.&.&.&.&.&. \cr 0&.&.&.&1&0&0 \cr .&.&.&.&0&cos \phi_2&e^{i\psi_2}
sin \phi_2 \cr 0&0&.&.&0&-e^{-i\psi_2}sin \phi_2&cos \phi_2 \cr
0&.&.&.&0&0&1 } \right )
\end{equation}
and the similar calculation of $\psi'_2$, $\phi'_2$. Generally one
should make $N-1$ steps in order to annulate $N-1$ elements of the
$|A>$. It is easy to see that each step of the ``squeezing" is action of the
unitary gate from $SU(2)$ acting on a qubit intentionally chosen in some complex direction.    Therefore one of the possible set of gates transformation of arbitrary state vector
to first vector of the ``standard" basis is established.

It is clear that these transformations of the isotropy group of $|1>$  are not optimal since they act at the first $N-2$ steps as ``zigzag" motions for alignment of $z|1>$  and $|A>$ along the geodesic in $CP(N-1)$. Only the final step belongs to the coset transformation
are optimal since drags almost ``squeezed" state vector to $z|1>$ along geodesic of $CP(1) \subset CP(N-1)$ \cite{Le1,Le2}.

\textbf{Theorem 2.}
It is possible to connect two normalized state vectors
by one unitary coset transformation generating motion along $CP(N-1)$ geodesic.

\textbf{Proof.}
It is enough to prove this statement for two states, say, now for $|1>$ and $|B>$.

Let me write the normalized state vector $|B>$ in the local coordinates
\begin{equation}\label{6}
\pi^i_{(j)}=\cases{\frac{b^i}{b^j},&if $ 1 \leq i < j$ \cr
\frac{b^{i+1}}{b^j}&if $j \leq i < N-1$},
\end{equation}
as follows:
\begin{eqnarray}\label{7}
|B>  =\sum_0^{N-1}
b^a(\pi^1_{j},...,\pi^{N-1}_{j})|a>,
\end{eqnarray}
where $\sum_{a=0}^{N-1} |b^a|^2= R^2$, and
\begin{eqnarray}\label{8}
b^0(\pi^1_{j},...,\pi^{N-1}_{j})=\frac{R^2}{\sqrt{R^2+
\sum_{s=1}^{N-1}|\pi^s_{j}|^2}}.
\end{eqnarray}
For $1\leq i\leq N-1$ one has
\begin{eqnarray}\label{9}
b^i(\pi^1_{j},...,\pi^{N-1}_{j})=\frac{R
\pi^i_{j}}{\sqrt{R^2+\sum_{s=1}^{N-1}|\pi^s_{j}|^2}},
\end{eqnarray}
i.e. $CP(N-1)$ is embedded in the Hilbert space ${\cal{H}}=C^N$.
Hereafter I will suppose $R=1$ and $j=0$.
The real measurement assumes some interaction of the measurement
device and incoming state. If we assume for simplicity that incoming
state is $|1>$ then all its isotropy group transformations arose from its
$H$-subalgebra will leave it intact.
The time-independent Hamiltonian $\hat{H}_B$ may be represented by the re-scaled ``quantum force" matrix
\begin{eqnarray}\label{10}
\hat{F} = \frac{1}{\hbar}\hat{H}_B= \left( \matrix{0& f^{1*} & f^{2*} &.&.& f^{N-1*} \cr
f^1 & 0 & 0 & 0 & 0 & 0                                       \cr
f^2&.&.&.&. &0\cr
.&.&.&.&. &0\cr
.&.&.&.&. &0\cr
f^{N-1} & 0 &.&.&. &0} \right),
\end{eqnarray}
generating coset unitary transformations represented by the matrix
\begin{eqnarray}\label{11}
 & \hat{T}(\tau,g) = \exp(i\hat{F} \tau) \cr =& \left( \matrix{\cos g\tau &\frac{-f^{1*}}{g} \sin
g\tau&\frac{-f^{2*}}{g}\sin g\tau &.&\frac{-f^{N-1*}}{g}\sin g\tau \cr
\frac{f^1}{g} \sin g\tau &1+[\frac{|f^1|}{g}]^2 (\cos g\tau -1)&[\frac{f^1
f^{2*}}{g}]^2 (\cos g\tau -1)&.&[\frac{f^1 p^{N-1*}}{g}]^2 (\cos g\tau -1)
\cr .&.&.&.&. \cr .&.&.&.&. \cr .&.&.&.&. \cr \frac{f^{N-1}}{g}\sin
g\tau &[\frac{f^{1*} p^{N-1}}{g}]^2 (\cos g\tau
-1)&.&.&1+[\frac{|f^{N-1}|}{g}]^2 (\cos g\tau -1)} \right),
\end{eqnarray}
where $g=\sqrt{\sum_{s=1}^{N-1}|f^s|^2}$ will effectively to
variate the incoming state $|1>$ dragging it along one of the geodesic in
$CP(N-1)$ toward final state (\ref{7}) \cite{Le1,Le3}. This matrix describe the process of the transition from one pure state to another, in particular between two
states connected by the geodesic
\begin{eqnarray}\label{12}
\hat{T}(\tau,g)|1>= \left( \matrix {\cos g\tau \cr \frac{f^{1}}{g} \sin
g\tau \cr \frac{f^{2}}{g} \sin
g\tau \cr . \cr . \cr . \cr \frac{f^{N-1}}{g}\sin
g\tau }
\right).
\end{eqnarray}
This vector being compared with the vector (\ref{7}) gives after simple algebra following solution:
\begin{eqnarray}\label{13}
f^i &=& \frac{W}{\hbar} \pi^i \\
g &=& \sqrt{\sum_{s=1}^{N-1}|f^s|^2} = \frac{W}{\hbar} \sqrt{\sum_{s=1}^{N-1}|\pi^s|^2} \\
\tau &=& \frac{\hbar}{W \sqrt{\sum_{s=1}^{N-1}|\pi^s|^2}} \arccos{ \frac{1}{\sqrt{1+\sum_{s=1}^{N-1}|\pi^s|^2}}}.
\end{eqnarray}
This result solves in fact the ``decision problem" for string of any length $N$ by acceptable ``cost"
$C=g\tau \leq \pi/2$ (see Figure 1) and polynomial time $T+\tau$, where $T \propto N$ is time of the squeezing procedure.  Note, that $\tau$ even decreases with the rise of the ``polarization" $x = \sqrt{\sum_{s=1}^{N-1}|\pi^s|^2}$ above some threshold $x \approx 1.39$, that so $T_{max}(1.39) \approx 0.8 \frac{\hbar}{W}$ where $W$ is energetic scale characterizing all dynamics (see Figure 2).

\begin{figure}[h]
  \includegraphics[width=2in]{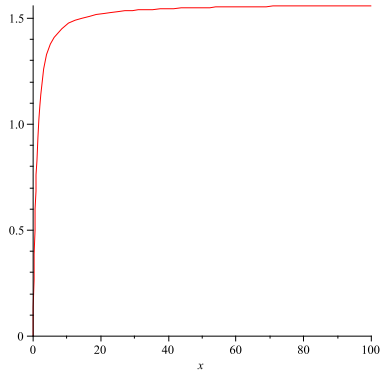}\\
  \caption{The cost required to reach target state along $CP(N-1)$ geodesic as function of polarization $x = \sqrt{\sum_{s=1}^{N-1}|\pi^s|^2}$}\label{fig.1}
\end{figure}

\begin{figure}[h]
  \includegraphics[width=2in]{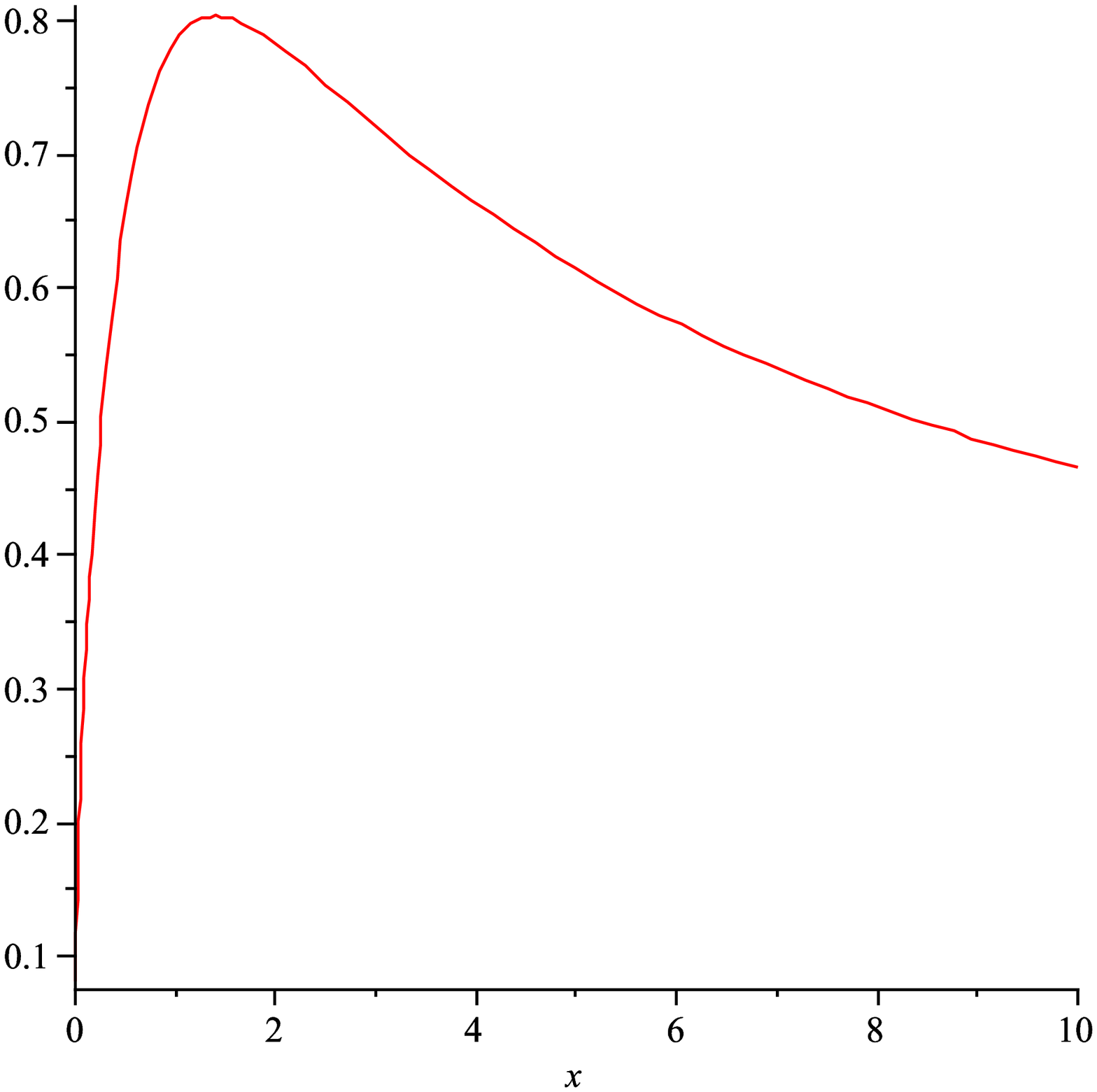}\\
  \caption{The time required to reach target state along $CP(N-1)$ geodesic as function of polarization $x = \sqrt{\sum_{s=1}^{N-1}|\pi^s|^2}$}\label{fig.2}
\end{figure}

\section{Local dynamical variables represented by vector fields}
The algorithm proposed above concerns the control by time-independent but  final-state-dependent Hamiltonian $\hat{H}_B$, since (\ref{10}) with $f^i = \frac{W}{\hbar} \pi^i$ should be used for manipulation of the register state. It is applicable to any  initial and target states $|A>$ and $|B>$. The example is, say, Quantum Fourier Transform $FT_M$, connected initial state $f=\sum_{x\in Z_M} f(x) |x>$ and $\hat{f}=\sum_{x\in Z_M} \hat{f}(x) |x>$
where $\hat{f}(x)=\frac{1}{\sqrt{M}}\sum_{y\in Z_M} f(y) \omega^{xy}$. Similar examples give us the actions of such unitary gates as Hadamar, phase shift and controlled-U gate whose final state may be connected by $CP(1)$ geodesic with an initial state.

The method of control given above is applicable in the semi-classical regime when the reaction of quantum state on the external field and inertial properties
of quantum state are negligible. But when uniliteral interaction is non-realistic and stiff control is not acceptable, arises more complicated problem of the dynamical control with self-consistent interaction. Dynamical quantum control is nothing but self-interaction
in combined system ``quantum state + control field".

Let me return now to the ``quantum control problem" \cite{MAN} as synthesis
of given n-qubit unitary $U \in SU(N)$ by the application of some time dependent Hamiltonian $\hat{H}(t) \in AlgSU(N)$ during time $T$,

\begin{equation}\label{16}
    \hat{H}(\tau) = \hbar \sum_{\alpha = 1}^{N^2-1}\Omega^{\alpha}(\tau) \hat{\lambda}_{\alpha}.
\end{equation}

The synthesis of unitary matrix $U \in SU(N)$ in formulation of \cite{MAN} from time dependent Hamiltonian $\hat{H}(t) \in AlgSU(N)$ during time $T$ is very complicated
since has not generally exact solution and requires exponential time to achieve acceptable accuracy. I propose to solve more ``flexible" problem: to find not only time-dependent but state-dependent Hamiltonian vector field on $CP(N-1)$. It correspond to the problem of  affine gauge field formulated in \cite{Le4,Le5,Le6}.

Interaction of any ``filter" $\hat{H}$ with incoming quantum state $|\Psi>$ depends only
on their relative ``orientation". It is clearly explained on the example of incoming polarization state of photon and the orientation of the $\lambda/4$ plate in \cite{Le4}.
Furthermore, only relative phases and amplitudes of photons have a physical sense for
their interaction with $\lambda/4$ plate. One may assume that it is correct and in general case of quantum interaction. It means that the filter action depends only upon the local coordinates (\ref{6}) in the complex projective Hilbert space $CP(N-1)$. The transition to the local coordinates is in fact non-linear \textbf{dynamical mapping}
onto $CP(N-1)$. \textbf{Small relative re-orientation of the filter and incoming state leads to small variation of  outgoing state.} This is a key point of all construction invoking to life the concept of the local dynamical variables (LDV) expressed by tangent vectors fields to $CP(N-1)$. It is convenient to rely upon ``orientation" of the filter given by the set of $N^2-1$  unitary group field parameters $\Omega^{\alpha}$ somehow related to space-time coordinates (I would like to note, that it is not so trivial problem as thought before). Then, since any state $|\Psi>$ has the isotropy group
$H=U(1)\times U(N)$, only the coset transformations $G/H=SU(N)/S[U(1)
\times U(N-1)]=CP(N-1)$ effectively act in $C^N$. Therefore the
ray representation of $SU(N)$ in $C^N$, in particular, the embedding
of $H$ and $G/H$ in $G$, is a state-dependent parametrization.
Technically the local $SU(N)$ unitary classification of the quantum motions requires the transition from the matrices of Pauli $\hat{\sigma}_{\alpha},(\alpha=1,...,3)$, Gell-Mann $\hat{\lambda}_{\alpha},(\alpha=1,...,8)$, and in general $N \times N$ matrices $\hat{\Lambda}_{\alpha}(N),(\alpha=1,...,N^2-1)$ of $AlgSU(N)$ to the tangent vector fields to $CP(N-1)$ in local coordinates \cite{Le1}.
Hence, there is a diffeomorphism between the space of the rays
marked by the local coordinates (\ref{6}) in the map
 $U_j:\{|\Psi>,|\Psi^j| \neq 0 \}, j\geq 0$
and the group manifold of the coset transformations
$G/H=SU(N)/S[U(1) \times U(N-1)]=CP(N-1)$ and the isotropy group of the corresponding ray.
This diffeomorphism is provided by the coefficient functions
\begin{equation}\label{17}
\Phi_{\sigma}^i = \lim_{\epsilon \to 0} \epsilon^{-1}
\biggl\{\frac{[\exp(i\epsilon \hat{\lambda}_{\sigma})]_m^i \Psi^m}{[\exp(i
\epsilon \hat{\lambda}_{\sigma})]_m^j \Psi^m }-\frac{\Psi^i}{\Psi^j} \biggr\}=
\lim_{\epsilon \to 0} \epsilon^{-1} \{ \pi^i(\epsilon
\hat{\lambda}_{\sigma}) -\pi^i \}
\end{equation}
of the local generators
\begin{equation}\label{18}
\overrightarrow{D}_{\sigma}=\Phi_{\sigma}^i \frac{\partial}{\partial \pi^i} + c.c.
\end{equation}
comprise of non-holonomic overloaded basis of $CP(N-1)$ \cite{Le1}.
This maps the unitary group $SU(N)$ onto the base manifold $CP(N-1)$. Now one may introduce Hamiltonian vector field as a tangent vector fields
\begin{equation}\label{19}
\overrightarrow{H}=\hbar \sum_{\alpha = 1}^{N^2 -1}\Omega^{\sigma}(\tau)D_{\sigma}=\hbar \sum_{\alpha = 1}^{N^2-1}\Omega^{\sigma}(\tau)\Phi_{\sigma}^i \frac{\partial}{\partial \pi^i} + c.c.
\end{equation}
whose control functions $\Omega^{\sigma}(\tau)$ may be found under the condition of self-conservation expressed as affine parallel transport of Hamiltonian vector field
$H^i=\Omega^{\sigma}(\tau)\Phi_{\sigma}^i$ agrees with Fubini-Study metric
\begin{equation}\label{20}
\frac{\delta (\Omega^{\sigma}(\tau)\Phi_{\sigma}^i) }{\delta \tau}= 0.
\end{equation}
The problem of finding ``control functions" $\Omega^{\sigma}(\tau)$ treated in the context of gauge field application as surrounding fields of quantum lump was discussed in \cite{Le5,Le6}. This approach requires future investigation and development in respect with signals propagation in quantum circuits too.

\section{Summary}
It is shown that two arbitrary normalizable vectors in $C^N$ may be connected by series of unitary transformation in polynomial time $T \propto N +\tau$ with acceptable ``cost"
$C=g\tau \leq \pi/2$.

\end{document}